# Electro-Thermal Characterization of Si-Ge HBTs with Pulse Measurement and Transient Simulation

Amit Kumar Sahoo, Sébastien Fregonese, Mario Weiß, Nathalie Malbert and Thomas Zimmer
Laboratoire IMS, CNRS - UMR 5218, Université de Bordeaux 1
Cours de la Liberation - 33405 Talence Cedex, France
Email : amit-kumar.sahoo@ims-bordeaux.fr

*Abstract*—This paper describes a new and simple approach to accurately characterize the transient self-heating effect in Si-Ge Heterojunction Bipolar Transistors (HBTs), based on pulse measurements and verified through transient electro-thermal simulations. The measurements have been carried out over pulses applied at Base and Collector terminals simultaneously and the time response of Collector current increase due to self-heating effect are obtained. Compared to previous approach, a complete calibration has been performed including all the passive elements such as coaxial cables, connectors and bias network. Furthermore, time domain junction temperature variations, current of heat flux and lattice temperature distribution have been obtained numerically by means of 3D electro-thermal device simulations. The thermal parameters extracted from measurements using HiCuM HBT compact model have been verified with the parameters extracted from electro-thermal transient simulation. It has been shown that, the standard R-C thermal network is not sufficient to accurately model the thermal spreading behavior and therefore a recursive network has been employed which is more physical and suitable for transient electro-thermal modeling.

## I. INTRODUCTION

The down scaling of electronic device dimensions has been expected to be the main way to continue miniaturization as seen by the International Technology Roadmap for Semiconductors (ITRS) to achieve better high frequency performance. Over the past decades this downscaling has allowed the semiconductor industry to gain significant progress in high speed and new circuit applications such as automotive radar, 100Gb/s data transfer etc. To achieve higher frequency performance, the transistor is made in such a way that its operating quiescent point is shifted to higher current densities and therefore power consumption of devices has increased significantly, resulting in self-heating effect in power and RF hetero-junction bipolar devices. Thermal issue is one of the key factors limiting the performance and reliability of the devices and integrated circuits, therefore systematic characterization of thermal effects inside the devices remains very important criteria to explore.

In general, self-heating effect characterization is based on steady state condition to extract thermal resistance and transient condition to extract thermal capacitance. A number of attempts have been accomplished to study self-heating effect in steady state conditions which are in general based on DC measurements at different ambient temperatures [1]. The transient temperature response can be obtained by pulses applied to the collector node and keeping the base current constant [2]. Measuring Base-Emitter voltage permits to determine the change in temperature inside the device. Although there are major limitations in this method because of difficulty in calibrations, like the passive elements related to coaxial cables and connectors, must be taken into account and should be accurately characterized. Various analytical models have been developed to investigate self-heating effect by connecting different electro-thermal networks like Foster network, Nodal network, Recursive network etc. at temperature node of the compact model [2]. In time domain, heat diffusion can be modeled in physical way by distributed electro-thermal network for pyramidal heat flow [3] or spherical heat flow [4]. A systematic study on comparison among different electro-thermal networks in the frequency domain has been proposed by [5]. It has been seen that, Recursive electro-thermal network provides the best compromise between accuracy, number of model parameters and physical basement.

In the first part of this paper, we discuss the electro-thermal transient simulation, where a pulse of electric power has been applied at Base-Collector junction of the device to obtain the transient response of temperature variation. In second part we discuss transient measurements, applying voltage pulses at Base and Collector terminal simultaneously and measuring the transient response of Collector current increase due to self-heating. The measurements have been calibrated taking all the passive elements related to coaxial cables, connectors, bias network etc. into account. In third part we have discussed electro-thermal modeling using HiCuM [6] compact HBT model and its limitations. The Recursive network proposed in [5], has been modified to apply in time domain case and has been verified with measurements and numerical simulation.

## II. ELECTRO-THERMAL TRANSIENT SIMULATIONS

3D electro-thermal transient simulations have been performed using Sentaurus Device Simulator taking ½ of devices placed on a semi-infinite Si-block with similar thickness as wafer one (300 µm) in order to validate measurement results. We consider the heat diffusion from Base-Collector junction to Si material (lower part of device) of

This research is funded in part by IMS Laboratory and University of Bordeaux 1.

device with STMicroelectronics BiCMOS9MW technology. The device under analysis has been made by Sentaurus Structure Editor (see Fig. 2) with key technological and geometrical parameters, $D_{ST}$ (height of shallow trench) = 0.3μm, $D_{DT}$ (height of deep trench) = 2.5 μm. Two thermodes have been placed at thermal node and thermal ground of the device respectively. A comparative study on thermal resistance for different device structure (from basic device taking only lower part and Deep Trench and then adding Shallow Trench, thick $SiO_2$ layer, SiGe Base, Polysilicon Base, floating metal layer etc.) has been demonstrated in the DOTFIVE report [7]. It has been seen that nearly about 5% of heat flux can diffuse through the upper part; therefore upper part has been neglected for simplicity in transient simulation. A pulse of power has been applied at thermode and transient response of temperature change has been calculated numerically as shown in the Fig. 1. The Fig. 2 presents temperature and heat flux distribution inside the device.

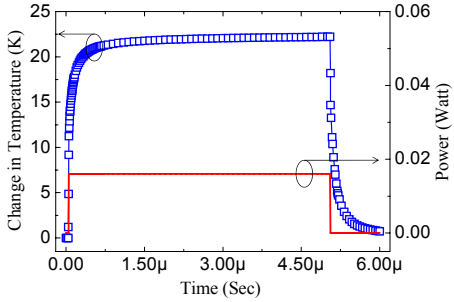

Fig. 1: Numerically simulated transient response of device temperature, when a Pulse (0.008Watt, 5000ns) of power is applied at thermode. Dimension of heat source ($L_E$ x $W_E$) =14.88 x 0.15 μm$^2$

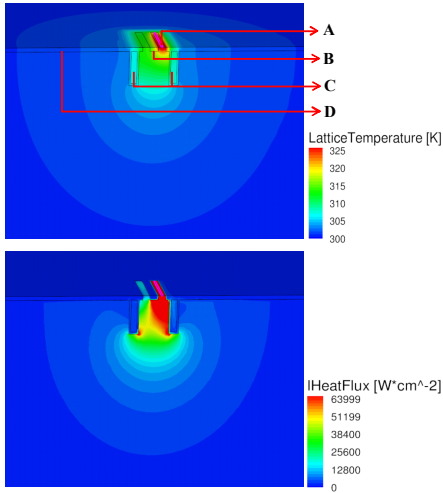

Fig. 2: Lattice Temperature and Heat Flux distribution inside the device at pulse time 5000 ns (see Fig. 1), [A: Heat source, B: Shallow Trench, C: Deep Trench, D: thick $SiO_2$ layer]. Dimension of the heat source ($L_E$ x $W_E$ =14.88 x 0.15 μm$^2$) is same as effective heated area (Base-Collector junction area) of the transistor, which is equivalent to the measured transistor with area of Emitter window ($L_E$ x $W_E$) = 15 x 0.27 μm$^2$.

### III. TRANSIENT MEASUREMENTS

On chip pulse measurements have been achieved with different Base-Collector pulse widths and bias conditions using MC2 Technology APMS LPM1/HPM1 pulse generator and Agilent 6633B system DC power supply on RF HBTs in Ground-Signal-Ground (GSG) configuration at room temperature. The measured NPN transistors are fabricated within the ST Microelectronics Si-Ge BiCMOS9MW technology [8], which is a quasi self-aligned trench isolated technology. The key figures of this technology are: current gain ($β$) = 950, transition frequency ($f_T$) = 230 GHz and maximum oscillation frequency ($f_{max}$) = 290 GHz.

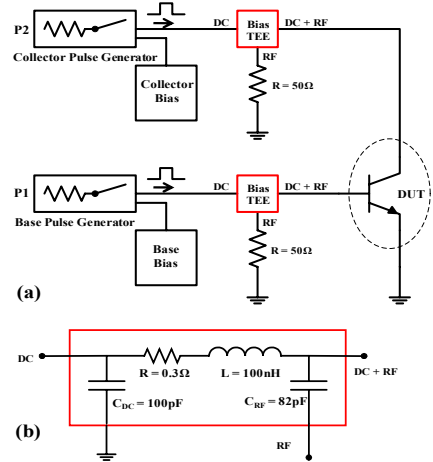

Fig. 3: Experimental setup (a) for pulse measurements, where pulses are applied through the DC port of the Bias network (b) and the RF port is grounded through 50 Ω resistances.

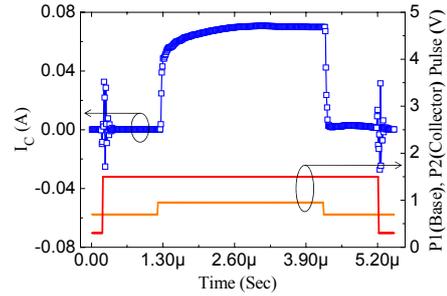

Fig. 4: Applied Base (P1=0.95Volt, 3000 ns) and Collector (P2=1.5Volt, 5000 ns) pulse and measured transient response of collector current for a DUT with dimension of Emitter window ($L_E$ x $W_E$) =15 x 0.27 μm$^2$.

Experimental setup for pulse measurements is described in Fig. 3. Pulses are applied through the DC port of the bias network where RF port is grounded with 50Ω resistance. We have several passive elements between pulse generator and device terminal arise due to internal resistance of pulse generator, bias network, co-axial cables and co-axial connectors. It has been seen through measurements that the capacitance ($C_{cab}$) and inductance ($L_{cab}$) of cables are most responsible for the noise arising in Base-Collector pulse profiles and the values of $C_{cab}$ and $L_{cab}$ increase linearly with cable length. So it should be necessary to optimize for accurate measurements and thermal modeling. Coaxial cables-connectors are measured by HP-4194A auto-balancing bridge type Impedance analyzer and the 16047E test fixture with open-short method. The length of the cables has also been optimized to minimize noise. Time domain Collector current has been measured as shown in (Fig. 4).

## IV. ELECTRO-THERMAL COMPACT MODELING

In this section thermal parameters for two different electro-thermal networks are extracted from transient measurement using compact model (Fig. 5) as well as transient electro-thermal numerical simulations.

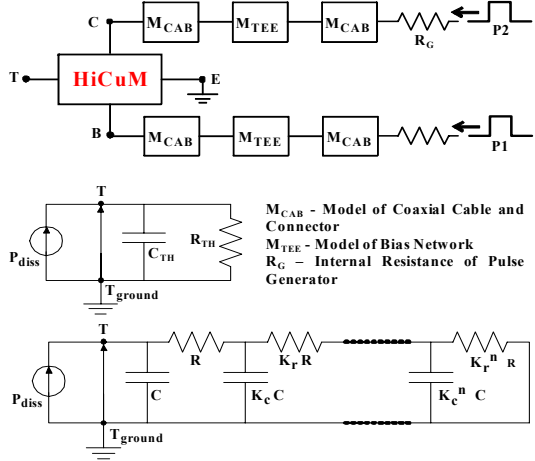

Fig. 5: Compact model simulation of the pulsed measurement configuration where electro thermal networks (two different networks, single $R_{TH}$-$C_{TH}$ and Recursive electro-thermal, have been utilized for transient thermal behavior comparison) have been connected at the temperature node of the HiCuM model

Circuit simulations were performed with Agilent ICCAP using two pulse sources that were connected through a model of passive elements to the base and the collector of the transistor model (Fig. 5). The model of passive elements includes Bias networks, coaxial cables and connectors of the measurement configuration are described in part III of this study. The internal network of the Bias-TEEs has been provided by MC2 and the passive elements of cables and connectors were extracted from measurements. BiCMOS9MW transistor models (CBE configuration) for 4 different emitter lengths were generated by a scalable HiCuM library provided by XMOD Technologies. The thermal node of the transistor model is either connected with the standard electro-thermal network including single $R_{TH}$ and $C_{TH}$ or with the Recursive network shown in Fig. 5. The recursive network is a physical approach to characterize the thermal spreading impedance of the transistor. The heat diffusion from the active region can be simplified as a spherical heat conduction path [4]. This also shows that every element is equivalent to a heat generating sphere. With further distance from the active region the area of a sphere (equithermal surface) increases. An elementary spherical slice at a distance z (towards thermal ground from heat source) with thickness $\Delta z$ is characterized by a capacitance $C(z)\Delta z$ and resistance $R(z)\Delta z$ [3] as follows,

$$R(z)\Delta z = \frac{1}{kA(z)}\Delta z \quad \text{and} \quad C(z)\Delta z = C_v A(z)\Delta z \quad (1)$$

Here k is the thermal conductivity of Si, $C_v$ is the specific heat at constant volume of Si and $A(z)$ is the area of the spherical equithermal surface element at a distance z. Therefore the time-dependent diffusion of heat can be modeled by a distributed Recursive network where, the respective thermal resistance element decreases and the capacitance element increases, which has been modeled by the multiplication factor $K_r$ (<1) and $K_c$ (>1) with R and C respectively (Fig. 5).

The thermal parameters for both networks are extracted from measured transient response of Collector current. A comparison between compact model simulation and measurements is shown in Fig. 6. Short and long pulse time extraction has been performed with single $R_{TH}$-$C_{TH}$ network which gives the minimum and maximum value of $C_{TH}$. The mid time pulse has been fitted with average value of $C_{TH}$. The same extraction has been performed with transient simulation as shown in Fig. 7. A time delay between measurement and compact model simulation using recursive network extracted without calibration has been observed in Fig. 6. Calibrated simulation including all passive elements is mandatory to achieve an accurate extraction.

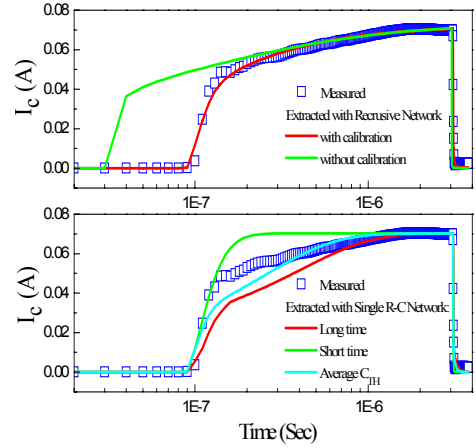

Fig. 6: Measured transient response of collector current (taken from Fig. 4) and extracted with HiCuM HBT model applying electro-thermal network at temperature node, time scale has been sifted to compare with simulation

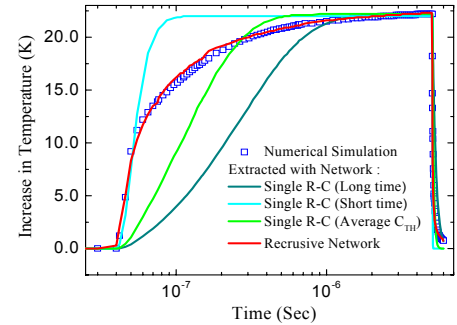

Fig. 7: Numerical simulations of time domain temperature change (taken from Fig. 1) of device and extracted with electro-thermal network

The measurements presented in Fig. 6 are very difficult to achieve due to the complex calibration process explained in section III. For validation purpose, transient electro-thermal simulation has been performed as described in section II. A power pulse has been applied at the Base-Collector junction in order to simulate time domain temperature response. The simulated transient temperature has been modeled with both electro-thermal networks as shown in Fig. 7.

The transient variation of the lattice temperature ΔT is mainly defined by the thermal capacitance whereas the steady state value is modeled by the overall thermal resistance, $R_{TH}=\Sigma R * k_r^i$, where i=0, 1, 2,.. n; the number of cell of the Recursive electro-thermal network. The components of the standard thermal network and of the recursive network, derived from electro-thermal modeling of numerically simulated transient temperature and measured collector pulse, are shown in Fig. 8 and 9.

## V. RESULTS

The variation of thermal resistance and thermal capacitance behavior with emitter area has been achieved from pulse measurements as well as transient simulation. A comparison of time domain thermal modeling between pulse measurements and numerical simulations with Single R-C (Fig. 8) and Recursive electro-thermal network with 9 R-C cells (Fig. 9) has been demonstrated.

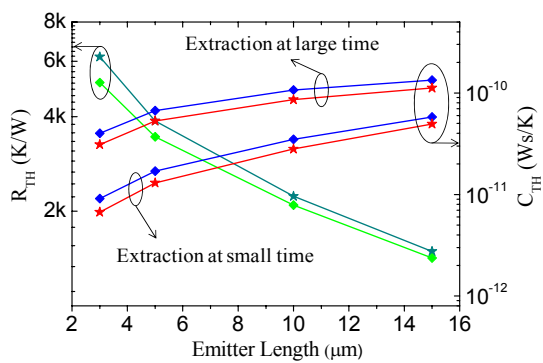

Fig. 8: Extracted Thermal Resistance and Capacitance form Measurements (diamonds) and Numerical simulations (stars) using Single electro-thermal network where $C_{TH}$ has been extracted at short time and long time pulse.

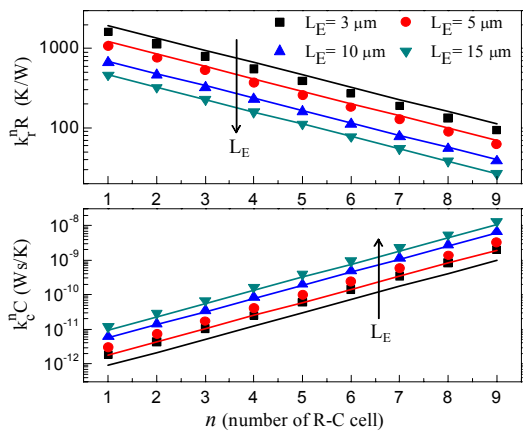

Fig. 9: Values for $R_{TH}$s ($k_r^n R$) and $C_{TH}$s ($k_c^n C$) of the different cells of Recursive network (according to Fig. 5) extracted from measurements (symbols) and simulation (lines) for an Emitter width $W_E=0.27\mu m^2$

From the comparison it is identified that fitting with recursive network is more accurate than using only single $R_{TH}$ and $C_{TH}$ (Fig. 6 and 7). This can be explained physically by thermal flux and lattice temperature distribution from heat source to thermal ground. According to the Fig. 2, the evolution of equithermal surface and equithermal volume element from heat source to ground can be represented by the decrease of thermal resistance and the increase of thermal capacitance as shown in Fig. 9. The lower value of $R_{TH}$ extracted from measurements can be caused by thermal flux through metal contacts that has not been taken into account for numerical device simulation.

## VI. CONCLUSION

In this study, the transient self-heating effect in Si-Ge HBTs, based on pulse measurements has been investigated. The extraction of $C_{TH}$ parameter in time domain for small device dimension can be achieved by employing a complete calibration. Transient device temperature variation has been achieved by means of 3D electro-thermal numerical simulations. Thermal parameters extracted from measurements have been verified with transient simulations for different Emitter dimensions. Two different electro-thermal networks have been compared to demonstrate time domain thermal spreading impedance. It is, therefore, evident that the modeling of transient thermal effect with the recursive network can achieve the extreme physical accuracy.

ACKNOWLEDGEMENT

This work is part of the Dotfive project supported by the European Commission through the Seventh Framework Program for Research and Technological Development. The Authors would like to thank XMOD TECHNOLOGIES for supplying the compact model parameters and STMicroelectronics for BiCMOS9MW wafer.